# Development and Optimization of a Multimedia Product

**Cristian Anghel, "Tibiscus" University of Timişoara, România**
**Vlad Muia, West University of Timişoara, România**
**Miodrag Stoianovici, C.E.O. Holedeck, Timişoara, România**

**ABSTRACT**.  This article presents a new concept of a multimedia interactive product. It is a multiuser versatile platform that can be used for different purposes. The first implementation of the platform is a multiplayer game called Texas Hold 'em, which is a very popular community card game. The paper shows the product's multimedia structure where Hardware and Software work together in creating a realistic feeling for the users.

**Introduction**

When we talk about multimedia & multi-user aplications most times we think about network based computer games. This is the most well known implementation of this concept. We can remember a lot of other aplications that are integrated in this category: multimedia presentations, video-conferences, computer-based training courses (CBTs) etc.

What's the meaning of Multimedia? **Multimedia** is media and content that utilizes a combination of different content forms. Multimedia includes a combination of text, audio, still images, animation, video, and interactivity content forms. Multimedia applications that allow users to actively participate instead of just sitting by as passive recipients of information are called *Interactive Multimedia.*

Multi-user, what is it? **Multi-user** is a term that defines a software application or operating system that allows concurrent access by multiple users of a computer. Time-sharing systems are multi-user systems.

Therefore, the two concepts are found mostly in applications which run on a computer network. These networks are composed of the Server which the main application runs on and the client computers from which the users have access to the application's interface. This solution generates high costs in terms of the hardware and Software licenses which have to be purchased for each PC.

In trying to find a cheaper solution we came up with the idea of using a single powerful machine (the equivalent of the server), with the possibility of interacting with more users simultaneously, as the best option. This is the way Holodeck One was born, a project witch is in the maturing and optimization process.

## 1 Holodeck One

Holodeck One is a versatile multimedia – multi-user system, which can be optimized and used for moast of the multimedia interactive applications in existence today. The forst application running on the system, and in the debug / optimization faze, is the Texas Hold 'em poker game.

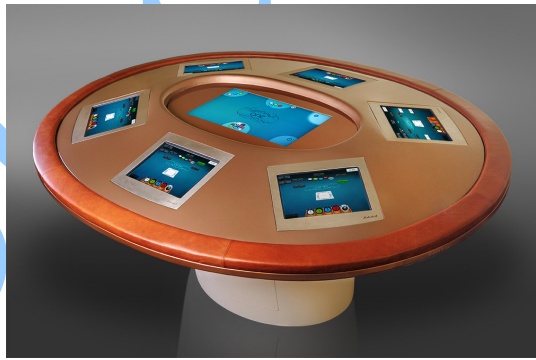

We will present the product in its two main components: Hardware and Software.

## 1.1 Holodeck One – Hardware

The main hardware parts of Holodeck One are: the oval table structure; 8 monitors and the computer that runs and manages the application.

The general shape of the system is that of an oval table with a central pillar. The general dimensions are: 220cm long, 175 cm wide, and 100 cm high.

This structure can accommodate up to 6 users simultaneously to which we can add, if needed, a system administrator. The administrators role is defining the functioning parameters of the application and overseeing it's evolution. The table's central pillar houses the computer system which runs the application and hides the connectors and cables which connect the system to the peripherals that interact directly with the users (monitors, speakers, touch screens). It is necessary that these cables be wary well grouped and organized so that they don't disturb the air flow inside the table, which cud lead to overheating of the system and the monitors.

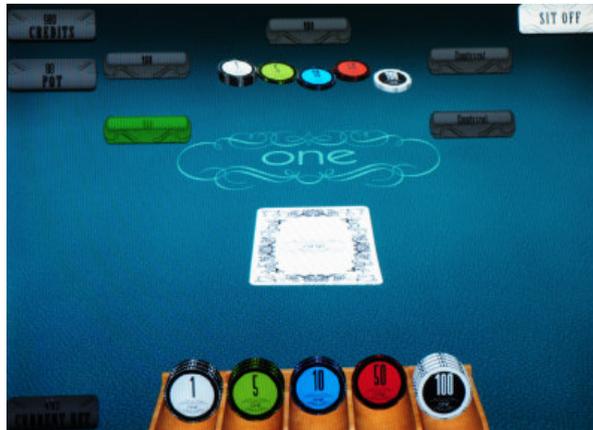

The main components of the system responsible for the interaction with the users are the 8 monitors. These monitors are TFT-LCD of various types. The central monitor is a 26 inch widescreen and generally shows the main interface of the application. Each user has in front of him/her a 15 inch 4:3 format monitor equipped with a touch screen panel. These touch screen devices are the method which allows the users to interact with the application. The monitor used by the administrator is a 7 inch touch screen. This one is mounted on a retractable arm and it can be 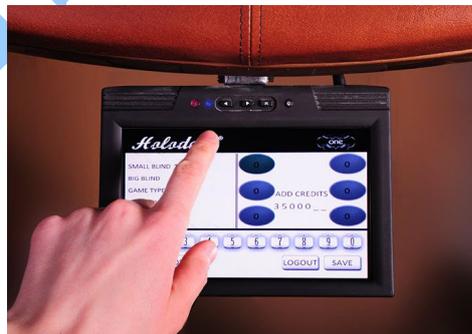 blocked, in this way preventing unauthorized access to the applications control menu. We can connect a keyboard and mouse to the system for debugging or testing the application, but when running the application, the touch screens are used for ensuring it's control.

The computer which animates the entire system must satisfy several conditions. It must ensure simultaneous and independent display on all 8 monitors, and so it uses the processing power of two video cards each capable of managing 4 displays each. The type and characteristics of the processor (CPU) and memory used will be chosen depending on the requirements of the application. In this perspective quad core CPU's can be used and up to 16 Gb of memory. The motherboard used must ensure optimal stability and reliability so it is checked to be 100% compatible with the components it houses. Taking this into consideration, only premium motherboards will be used from the top manufacturers. Extra precaution will also be taken in choosing the power source that will feed the whole system with electricity as it is a key element in ensuring the stability and reliability of the system.

Choosing it is first of all determined by the sum of consumption of the two graphic accelerators. The case witch houses this powerful and robust system was especially designed for it. It's main quality is ensuring the coolest possible functioning of the system for a minimum of noise pollution. For this purpose the motherboard is rotated to the right at a 90 degree angle to the ATX standard. Applying this method we use the natural heat convection phenomenon, and eliminating excess fans to vent the heat outward. A Positive Pressure Airflow is created and the warm air exits the case via it's top side in a natural fashion. After our calculations it's possible to 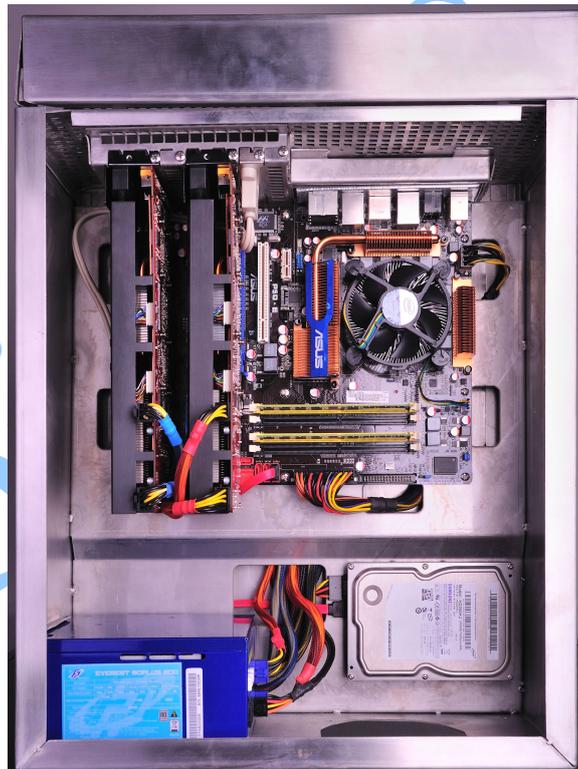 efficiently cool even the hottest systems using only 3 120 mm fans of medium rpm positioned in key places inside the case.

### 2.2 Holodeck One – Software

At first Texas Hold'em was chosen as the application to pot the system through it's paces. Hold 'em is the most popular poker game in the casinos and poker card rooms across North America and Europe, as well as online.

34

Hold 'em is a community card game where each player may use any combination of the five community cards and the player's own two hole cards to make a poker hand, in contrast to poker variants like stud or draw where each player holds a separate individual hand.

The application is in fact an assembly of 3 interconnected applications which use the TCP/IP port and the local IP address 127.0.0.1 for communicating. The software components are: the server, the client and the admin (the management application).

The Server is the main application of the product and it's 2 main roles are:
a) Processing of all data received from the other instances. After processing it is the one that decides what data will be sent on to the clients.
b) Display's on the central monitor the processed data.
c)

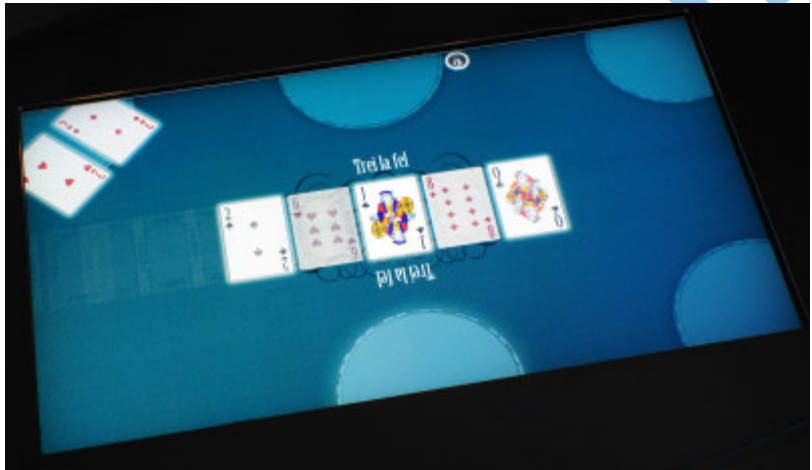

The client is responsible for receiving commands from the product's users. 6 Client applications run simultaneously on the table. These use RAW IMPUT to receive input from more users simultaneously. On each action upon a Client, this sends data to the Server to be processed.

The Admin is the application which changes and checks the products settings. In order for it to be used it is necessary to introduce a 6 digit PIN code. Any setup or check of the product is done through this application.

On startup the first application launched and running is the Server. It is the one which will launch the 6 clients successively and finally the Admin.

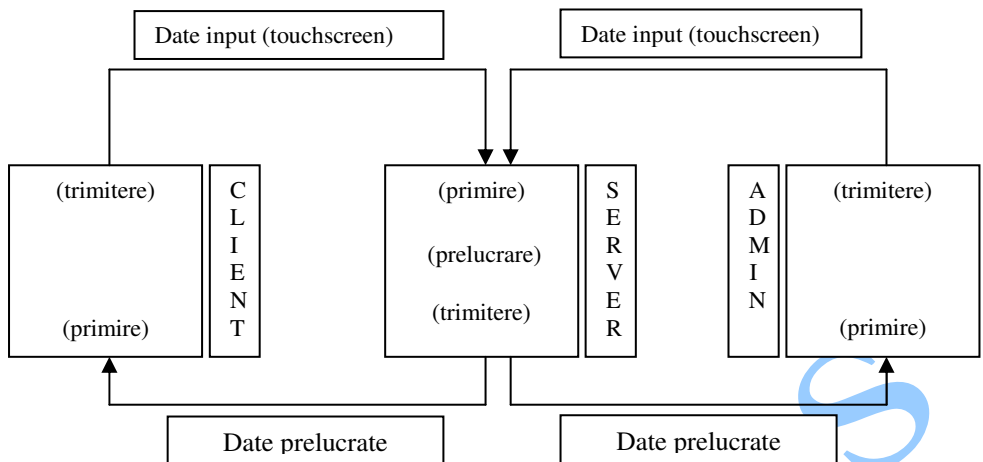

## Conclusion

Starting from the complex and costly solutions which depend on the existence of a computer network, Holodeck One differentiates itself as the efficient way of implementation for a multimedia – multi-user application. Holodeck One is perfect for most of these applications.

## References

http://en.wikipedia.org

http://msdn.microsoft.com/en-gb/default